\documentclass[12pt]{article}
\usepackage{graphicx}
\advance\textheight by \topskip
\begin{document}
\baselineskip 0.8cm
\thispagestyle{empty}
\begin{flushright}
KUCP0168\\
September. 21, 2000\\
\end{flushright}
\vskip 2 cm
\begin{center}
{\LARGE \bf Quantum Fluctuations of Bulk Inflaton \\
in Inflationary Brane World }
\vskip 1.7cm

{\bf Shinpei Kobayashi}
\footnote{E-mail: shinpei@phys.h.kyoto-u.ac.jp}
{\bf ,}
{\bf Kazuya Koyama}
\footnote{E-mail: kazuya@phys.h.kyoto-u.ac.jp}
{\bf and}
{\bf Jiro Soda}
\footnote{E-mail: jiro@phys.h.kyoto-u.ac.jp } \

\vskip 1.5mm

\vskip 2cm
 $^{1, 2}$ Graduate School of Human and Environment Studies, Kyoto
University,
       Kyoto  606-8501, Japan \\
 $^3$ Department of Fundamental Sciences, FIHS, Kyoto University,
       Kyoto, 606-8501, Japan \\
\end{center}

\vskip 1cm
{\centerline{\large\bf Abstract}}
\begin{quotation}
\vskip -0.4cm
The inflationary scenario for the brane world driven by the bulk
inflaton is proposed. The quantum fluctuations of the inflaton is 
calculated and compared to those of the conventional 4-dimensional 
inflationary scenario. It is shown that the deviation of the primordial 
spectrum of this model from that of the conventional one is too small 
to be observed even if $AdS$ radius is very large. 
Hence, it turns out that the inflation caused by the bulk inflaton
 is viable in the context of brane world cosmology. 
\end{quotation}

\newpage

\section{Introduction}
The possibility that our universe is a four-dimensional hypersurface 
(3-brane) embedded in the higher-dimensional spacetime 
was known before and forgotten for a long time \cite{Rubakov1,BW2}. 
Recently,  
stimulated by M-theory, this possibility comes to be argued eagerly.
In such a current, 'An alternative to compactification' was suggested 
by Randall and Sundrum (RS) \cite{RS}. 
It is fascinating because the internal dimension
is not compact and hence there is no stability problem.
The bulk of this model consists of two five-dimensional anti-de Sitter 
spacetimes ($AdS_5$).
On the brane, there are 0-mode and massive modes of graviton in the
linearlized gravity as in the Kaluza-Klein 
compactification. The Newtonian gravity on the brane is realized by 
0-mode \cite{GT,GKR}, on the other hand, massive modes give the
correction to Newtonian gravity . This is the outline of the RS model.

Since the RS model is proposed,  the cosmological consequences of the
model have been studied and found that it is completely consistent with 
the present cosmological observations \cite{S}-\cite{K}.
However the validity of this model in the context of the very early universe 
is not investigated completely. 
Especially, the detailed analysis of the primordial spectrum of the 
fluctuation during the inflation is lacked.

In the RS model, the tension is balanced with the bulk cosmological constant.
This has allowed the Minkowski spacetime as a ground state. In the
presence of the cosmological matter, the evolution of the spacetime can be
described by the Friedmann equation at low energy. If we assume the
scalar field as a cosmological matter, the inflationary solution will
 arise because the effective tension of the brane exceeds the bulk 
cosmological constant. This is a naive extension of the conventional 
inflationary scenario \cite{BWC0}-\cite{MWBH}.

But considering theories of particle physics (e.g.String Theory), 
many scalar fields live in the higher-dimensional spacetime (bulk).
So it is natural to consider inflatons in the bulk - $AdS_5$.
Suppose that a scalar field lives in the bulk and has a potential.
When it slow-rolls, what will happen?
The potential works as the effective cosmological constant 
and it makes non-zero Hubble constant $H$ on the brane.
In other words, the de Sitter expanding universe can be realized 
by changing the bulk cosmological constant.
Furthermore, this scalar field continues rolling down, oscillates around 
some vacuum and get to some stable state with the reheating process.
This model does have all properties which the conventional inflationary 
scenario has and it has the extra dimension.
The purpose of this letter is to study the nature of the quantum
fluctuations in this inflationary brane world. 
As a result, we will show that our model is observationally
acceptable as an inflationary scenario in the brane world cosmology.

The organization of this paper is as follows:
In Section 2 we shall present the 
inflation model driven by the bulk inflaton. In Section 3 we shall 
calculate the quantum fluctuations of the bulk inflaton. 
Final section is devoted to the conclusion. 

\section{Inflation Driven by the Bulk Inflaton}

How can we make the brane which expands exponentially?
One of the most popular and natural ideas to make the de Sitter brane 
is to put an inflaton on the brane. This model is not different from the 
ordinary four-dimensional inflationary scenario in respect of using the 
four-dimensional matter. 
The models using the tension of the brane also exist \cite{BWC0}-\cite{LMW}.
But our model is different from any other models proposed so far \cite{A}.
In our model, the de Sitter brane is realized by the potential of the 
scalar field in the bulk.
We see this mechanism in detail.

The action of this system can be divided
into three parts, 
\begin{equation}
S = S_{grav.} + S_{brane} + S_{scalar}\ . 
\end{equation}
The first term of the action is the usual Einstein-Hilbert
action with a negative cosmological constant:
\begin{equation}
S_{grav.} = \frac{1}{16\pi G_5} \int \,d^5 x \sqrt{-g_{5}}
                    \big(R -2\Lambda_{5} \big)\ ,
\end{equation}
where $g_{5}$ and $G_{5}$ are the five-dimensional metric and
gravitational constant, $R$ denotes the five-dimensional Ricci scalar.  
The five-dimensional cosmological constant $\Lambda_{5}$ is related to 
the $AdS$ radius $\bar{l}$ by the following equation,
\begin{equation}
\Lambda_5 = -\frac{6}{\bar{l}^2}\ .
\label{eq:lambda}
\end{equation}
Note that $\bar{l}$ is the $AdS$ radius without a scalar field.
The second term of the action is the brane action:
\begin{equation}
S_{brane} = -\sigma \int \,d^4 x \sqrt{-g}\ .
\end{equation}
Here $\sigma$ works as the tension of the brane. The last term of $S$
is the action of the scalar field living in the bulk:
\begin{equation}
S_{scalar} = -\int \,d^5 x \sqrt{-g_{5}}
               \Big(\frac{1}{2}(\nabla\phi)^2 + U(\phi) \Big)\ ,
\end{equation}
where $\nabla$ is the covariant derivative with respect to $g_{5}$, 
and $U(\phi)$
is the potential of the scalar field, which plays an important role.

Now turn to find the solution of this system.
If there is no scalar field, the geometry of the bulk is represented 
by $AdS_5$ and the metric of $AdS_5$ becomes 
\begin{equation}
ds^2 = dy^2 +(\bar{l}H)^2 \  
     \sinh^2 (y/\bar{l}) (-dt^2 +H^{-2} \exp{(2Ht)} \delta_{ij}dx^i dx^j)\ .
\label{eq:metric}
\end{equation}
Here $H$ is the four-dimensional Hubble constant and $y$ denotes the 
extra dimension. We consider a four-dimensional 
hypersurface (3-brane) at 
$y=y_0$ and the location of the brane is related with $H$ as follows, 
\begin{equation}
H = \frac{1}{\bar{l}\sinh{(y_{0}/\bar{l})}} .
\label{eq:branelocation}
\end{equation}
Following the Randall-Sundrum model, we glue this $AdS_5$ and the copy 
of it at $y=y_0$. So the bulk has $Z_2$ symmetry
along the $y$-direction. In this coordinate system, the range of $y$ is 
$0 \le y \le y_0$.
Calculating Einstein equations and junction conditions at $y=y_0$,
we obtain the condition for the Hubble constant
\begin{equation}
H^2 = \Big(\frac{4\pi G_5}{3} \sigma \Big)^2 + \frac{\Lambda_{5}}{6}\ .
\label{eq:Hubble}
\end{equation}
In the RS model, the tension is tuned to
\begin{equation}
\sigma = \frac{3}{4\pi G_{5}\bar{l}}\ .
\end{equation}
From (\ref{eq:lambda}), we can see that $H$ becomes zero, then 
the brane is flat.
But in this paper, we want to make the de Sitter expanding brane,
that is, the brane having the positive Hubble constant.

Here we have two ways to realize $H > 0$.
One of them is to increase $\sigma$ and another is to decrease
$|\Lambda_5|$. Our model is the latter.
We consider the potential $U(\phi)$ to decrease $|\Lambda_5|$.
This potential can be regarded as the effective cosmological constant, 
that is, we can define the effective cosmological constant $\Lambda_{eff}$,
\begin{equation}
\Lambda_{eff} \sim \Lambda_{5} + U(\phi)\ .
\end{equation}
Here we neglect some coefficients. Note that we consider the situation 
where the slow-roll conditions are satisfied, so we neglect the 
kinetic term of the scalar field. 
$\Lambda_{eff}$ is related to the new $AdS$ radius $l$ as 
\begin{equation}
\Lambda_{eff} = -\frac{6}{l^2}.
\end{equation}
With the change of $AdS$ radius, the metric changes as 
\begin{equation}
ds^2 = dy^2 +(lH)^2 \  
     \sinh^2 (y/l) (-dt^2 +H^{-2} \exp{(2Ht)} \delta_{ij}dx^i dx^j)\ .
\label{eq:newmetric}
\end{equation}
Furthermore, we can rewrite (\ref{eq:Hubble}) as follows
\begin{equation}
H^2 = \Big(\frac{4\pi G_5}{3}\sigma \Big)^2 + \frac{\Lambda_{eff}}{6}\ .
\end{equation}
We consider the following conditions 
\begin{equation}
\Lambda_{5} < \Lambda_{eff}  ,
\label{eq:twolambda}
\end{equation}
\begin{equation}
\Lambda_{eff} < 0 .
\label{eq:lambdaeff}
\end{equation}
Eq.(\ref{eq:twolambda}) is the condition to realize the de Sitter
brane and eq.(\ref{eq:lambdaeff}) is the condition to make the bulk 
$AdS_5$. Thus we have $AdS_5$ with the de Sitter brane. 
This is a novel inflationary scenario.
When a scalar field starts slow-rolling, observers living 
on the brane see inflation.
Our model is different from other models which treat 
the inflatons confined on the brane.
After the scalar field slow-rolls along the potential, it oscillates  
around the vacuum and gets to a stable state.
If we tune the minimum of the potential to the Randall-Sundrum value, 
the system finally
gets to the Randall-Sundrum system (flat brane scenario).

\section{Quantum Fluctuations}
 
 We want to investigate the fluctuations of the scalar field $\phi$ 
living in the bulk ($AdS_5$) and evaluate it on the brane. 
We consider the zero-th order of the slow-roll approximation, that 
is, $U(\phi)\sim const.$. Therefore the spacetime of the brane is 
exact de Sitter. 
The equation for the fluctuations is Klein-Gordon equation in $AdS_5$: 
\begin{equation} 
\left[\partial_y^2 +\frac{4}{l}\coth(y/l)\partial_y 
-\frac{1}{(lH)^2\sinh^2(y/l)}(\partial_t^2 +3H\partial_t)
+\frac{1}{l^2\sinh^2(y/l)e^{2Ht}}
\partial_{\mbox{\boldmath $x$}}^2 \right]\delta\phi=0 .
\end{equation}
As $AdS_5$ is highly-symmetric, we can use the separation of variables.
The fluctuation $\delta \phi$ can be expanded as 
\begin{equation}
\delta\phi(y,t,\mbox{\boldmath $x$})
= \int \,d^3 \mbox{\boldmath $p$} \,dm \ \ \left(a_{pm}f_{m}(y)g_{pm}(t)
   e^{i\mbox{\boldmath $px$}} + (h.c.)\right)\ . 
\label{eq:bunri}
\end{equation}
Here, $a_{pm}$ represents the annihilation operator of 
the field theory in $AdS_5$ 
and $a_{pm}$ satisfies usual commutation relation,
\begin{equation}
\Big[a_{pm} , a_{p^{\prime}m{\prime}}^{\dagger}\Big] = \delta (p-p^{\prime})
                                      \delta (m-m^{\prime})\ .
\end{equation}
The vacuum can be defined with $\{a_{pm}\}$ as 
\begin{equation}
a_{pm} |0> = 0 , \ \ for \ \forall \mbox{\boldmath $p$} , m .
\end{equation}
Now the equations for $f_m(y)$ and $g_{pm}(t)$ become
\begin{equation}
f_{m}^{\prime\prime} (y) +\frac{4}{l}f_{m}^{\prime}(y)\coth{(y/l)}
   + \frac{m^2}{(lH)^2 \sinh^2{(y/l)}}f_{m}(y)=0\ ,
\label{eq:f}
\end{equation}
\begin{equation}
\ddot{g}_{pm}(t)+3H \dot{g}_{pm}(t)
+ \Bigl(p^2 e^{-2Ht} + m^2 \Bigr)g_{pm}(t)=0\ ,
\label{eq:g}
\end{equation}
where prime denotes derivative with respect to $y$, dot denotes derivative 
with respect to $t$ and $m$ 
corresponds to the Kaluza-Klein mass which appears in the Kaluza-Klein 
compactification. 

As the background has $Z_2$ symmetry along the $y$-direction, it is 
natural to impose the $Z_2$ symmetry on the scalar field also. 
Then, the junction condition becomes
\begin{equation}
\partial_{y} \delta\phi(y,t,\mbox{\boldmath $x$})\mid_{y_0} =0\ .
\label{eq:junction} 
\end{equation}
So we solve the equations under Neumann boundary condition. As to 
the condition of $g(t)$, we choose the Bunch-Davis vacuum, which 
coincides with the Minkowski vacuum at high-frequency ($p/aH \to \infty$,
$a$ is a scale factor).

Under these conditions, we can solve (\ref{eq:f}) and (\ref{eq:g}) 
uniquely. We get 0-mode ($m=0$) and massive modes ($m>{3H}/2$). 
The massive modes with $0<m<3/2$ are not normalizable. 
So we exclude these modes 
from solutions. 
The solution for 0-mode is
\begin{eqnarray}
f_0(y) &=& \frac{1}{lH \sqrt{l\sinh(y_0/l)\cosh(y_0/l)-y_0}}\ ,\\
g_{p0}(t) &=&\frac{\sqrt{\pi}}{2}H(-\eta)^{3/2}
                          H_{3/2}^{(1)}(-p\eta)\ ,
\end{eqnarray}
where $H^{(1)}$ is the first Hankel function and $\eta$ is conformal 
time defined as 
\begin{equation}
\eta \equiv -e^{-Ht} .
\end{equation}
Similarly, the solutions for massive modes are
\begin{eqnarray}
f_m(y) &=& \frac{1}{lH\sqrt{l(\zeta(\beta)+\xi(\beta))}}
           \sinh^{-2}{(y/l)} \nonumber \\
       & & {}{}\ \ \ \times \Big(P_{-1/2 +i\beta}^{-2}(\coth(y/l))
       -\alpha(y_0/l)Q_{-1/2 +i\beta}^{-2}(\coth(y/l))\Big)\ , \\
g_{pm}(t) &=& \frac{\sqrt{\pi}}{2}H(-\eta)^{3/2}e^{-\pi \beta /2}
                          H_{i\beta}^{(1)}(-p\eta)\ ,
\end{eqnarray}
where
\begin{eqnarray}
\beta &=& \sqrt{\frac{m^2}{H^2}-\frac{9}{4}}\ , \\ 
\zeta(\beta) &=&
      \left| \frac{\Gamma(i\beta)}{\Gamma(5/2 +i\beta)}\right|^2\ ,\\
\xi(\beta) &=& \left| \frac{\Gamma(-i\beta)}{\Gamma(5/2 -i\beta)}
 -\pi \alpha(y_0/l)\frac{\Gamma(-3/2+i\beta)}{\Gamma(1+i\beta)}\right|^2\ ,
\end{eqnarray}
\begin{equation}
\alpha(y_0/l) = 
\frac{(5/2+i\beta)P_{1/2+i\beta}^{-2}(\coth(y_0/l))
+(3/2-i\beta)\coth(y_0/l)P_{-1/2+i\beta}^{-2}(\coth(y_0/l))}
              {(5/2+i\beta)Q_{1/2+i\beta}^{-2}(\coth(y_0/l))
+(3/2-i\beta)\coth(y_0/l)Q_{-1/2+i\beta}^{-2}(\coth(y_0/l))}\ .
\end{equation}

We can calculate the vacuum expectation value of $\delta\phi^2$ using  
these solutions, but note that $\delta\phi$ is the five-dimensional 
scalar field. The four-dimensional scalar field $\phi_4$ is related 
to $\phi$ by the following equation, 
\begin{equation}
\phi_4 = l^{1/2} \phi .
\end{equation}
Consequently, the four-dimensional vacuum expectation value of 
$\delta\phi_4^2$ on the brane becomes
\begin{eqnarray}
<\delta\phi_4(y_0,t,\mbox{\boldmath $x$})^2>
 &=& l<\delta\phi(y_0,t,\mbox{\boldmath $x$})^2> \nonumber \\
 &=& l\int \, d^3 \mbox{\boldmath $p$} \Big(|f_0|^2 |g_{p0}|^2
   + \int_{3/2}^{\infty} \,dm \ |f_m|^2 |g_{pm}|^2 \Big) \\
 &\equiv& \int \, d^3 \mbox{\boldmath $p$} P(\mbox{\boldmath $p$})
\end{eqnarray}
Here, we define $P(\mbox{\boldmath $p$})$ as the power spectrum. 
We are interested in the fluctuations on the brane, 
so we evaluate it at $y=y_0$. 
And we sum up $m$, since $m$ is the momentum of the extra
dimension. 
We can not see each $m$ but the summation.

Here is the point we have to notice. It is that in the large $m$ and 
$p$ limit, 
$|g_{pm}|^2$ behaves 
\begin{equation}
|g_{pm}|^2 \sim \frac{H^2}{2\sqrt{m^2 + (-p\eta)^2}}  .
\label{eq:div}
\end{equation}
Here we remove the conformal factor $(-\eta)^{3/2}$ to compare 
the field theory in Minkowski spacetime.
And Hubble constant $H$ disappears when $|g_{pm}|^2$ is multiplied 
by $|f_{pm}|^2$.
The equation (\ref{eq:div}) shows that the summation has the 
logarithmic divergence. This ultra-violet divergence appears in the 
five-dimensional field theory in Minkowski spacetime as well. 
For this reason, this divergence is not the real divergence in the
sense that we can not observe it. After all, we can conclude that we 
have to subtract this divergence to get the correct expectation value.

The subtraction of $|g_{pm}|^2$ can be carried out as 
\begin{equation}
\frac{|g_{pm}|^2}{(-\eta)^{3/2}}-\frac{H^2}{2\omega} .
\end{equation}
Here $\omega = \sqrt{(-p\eta)^2 + m^2}$ and we inserted $H$ for the reason 
as mentioned before.  
Thus subtracted $|g_{pm}|^2$ behaves as ${\cal O}(\frac{1}{m^3})$ 
for large $m$  
and, hence, no divergence arises in the integration with respect to $m$.
Finally we get the power spectrum of the physical expectation value as
\begin{equation}
P(\mbox{\boldmath $p$})
 = l\left( |f_0|^2 \left[|g_{p0}|^2 
-(-\eta)^3\frac{H^2}{2\sqrt{(-p\eta)^2}}\right]
              + \int_{3/2}^{\infty} \,dm \ |f_m|^2
       \left[|g_{pm}|^2-(-\eta)^3\frac{H^2}{2\omega}\right] \right)\ .
\end{equation}

Next, we analyze the $P(\mbox{\boldmath $p$})$ on the brane,
especially the effect of massive modes. The reason why we are
interested in massive modes is that massive modes appear due to 
the extra dimension. In the four-dimensional theory,
Kaluza-Klein mass does not actually exist. So it is massive modes 
that show the difference between the Einstein theory and the brane-world 
theory. 

\begin{figure}
\centerline{
\includegraphics[width=10cm]{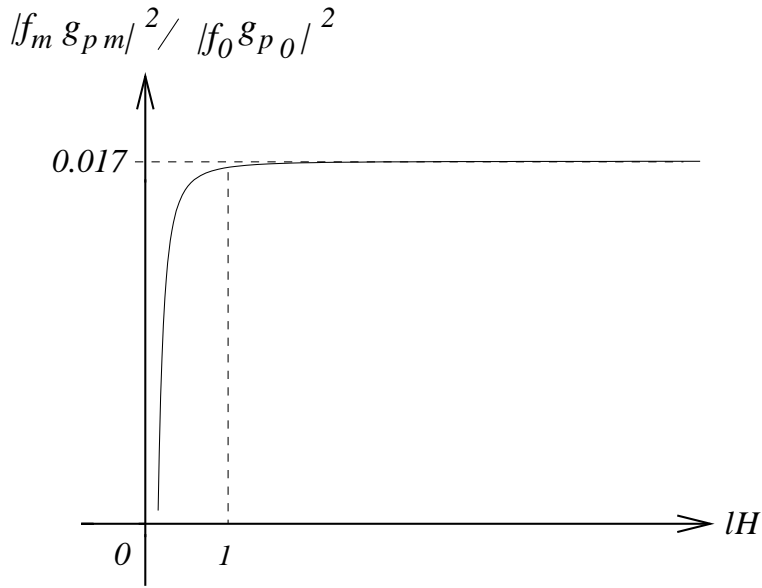}}
\caption{massive mode/0-mode ratio (evaluated at $m=1.9$)}
\label{fig:ratio}
\end{figure}

Fig.\ref{fig:ratio} is the $lH$-dependence of the ratio of massive modes 
amplitude to the 0-mode amplitude
(i.e.$|f_{m}g_{pm}|^2/|f_{0}g_{p0}|^2$).
\footnote{If we set $m=1.9$, $|f_m|^2/|f_0|^2 \sim 0.080$ and 
$|g_m|^2/|g_0|^2 \sim 0.21$. $|f_m g_m|^2$ takes the maximum value around 
$m=1.9$  so we use $m=1.9$.} \
We evaluate it at $-p\eta=1$ (horizon-crossing). 
As $l$ is the $AdS$ radius, it can be said that $l$ defines the scale 
of the compactification. In the RS model, the effect of massive modes 
dominates as $l$ becomes large. When the radius of the compactification 
becomes large, we come to be able to see the extra dimension and 
to feel the effect of the massive modes.
In our case, the same phenomenon occurs at $lH < 1$ .
But $lH \gg 1$, the ratio approaches to the constant value. 
Since the $lH$-dependence of the ratio is determined only by $f(y)$, 
we pay attention to the asymptotic form of $f(y)$ at large $lH$.
Using (\ref{eq:branelocation}), 
$|f_m|^2/|f_0|^2$ can be written as
\begin{equation}
\frac{|f_m(y_0)|^2}{|f_0(y_0)|^2} \stackrel{lH\to\infty}{\sim}
 \frac{2}{3\pi}\frac{m^2-\frac{9}{4}}{m^2}
\label{eq:asymptotic}
\end{equation}
This equation shows that if we determine the value of $m$, 
$|f_m|^2/|f_0|^2$ approaches to constant at large $lH$.

This result can be understood in the following way.
Following \cite{GS,LMW}, we rewrite (\ref{eq:f}) by 
using $\tilde{y}$ defined as 
\begin{equation}
\sinh{\tilde{y}} \equiv \sinh^{-1}{(y/l)} .
\end{equation}
In this relation we can see that $y \to 0$ corresponds to 
$\tilde{y} \to \infty$.
Setting $f(\tilde{y})=(\sinh{\tilde{y}})^{3/2} F(\tilde{y})$, 
the equation for $F(\tilde{y})$ becomes Schr\"{o}dinger-like form,
\begin{equation}
-\frac{d^2 F(\tilde{y})}{d\tilde{y}^2}+V(\tilde{y})F(\tilde{y})
=m^2 F(\tilde{y})
\end{equation}
where $V(\tilde{y})$ is so-called 'volcano' potential,
\begin{equation}
V(\tilde{y}) = 
\frac{15}{4\sinh{\tilde{y}}}+\frac{9}{4}
-3\coth{\tilde{y}_0}\delta(\tilde{y}-\tilde{y_0}).
\end{equation}

\begin{figure}
\centerline{
\includegraphics[width=12cm]{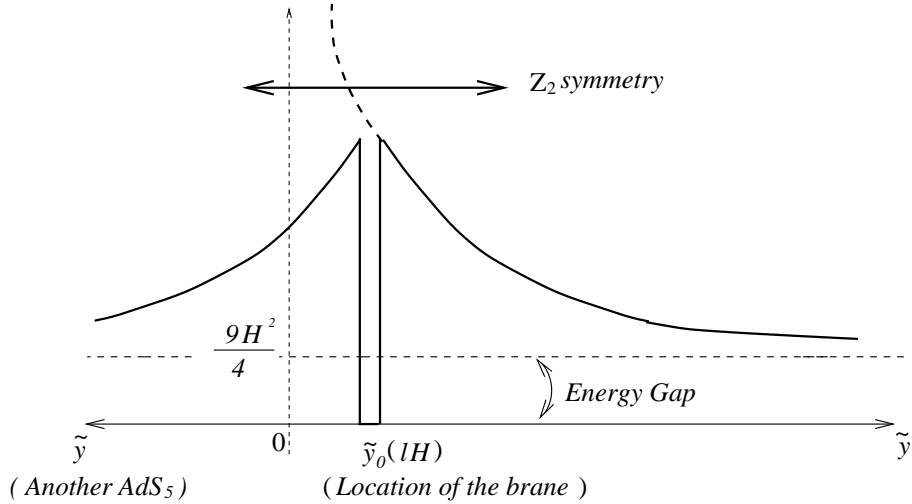}}
\caption{Volcano potential is shown. $\tilde{y_0}$ denotes the location 
of the brane and two $AdS_5$s are glued together at 
$\tilde{y}=\tilde{y_0}$. If $\tilde{y_0}$ becomes large (i.e. $lH$ 
becomes large), the height of the volcano becomes low.}
\label{fig:volcano}
\end{figure}

This potential is shown in fig.\ref{fig:volcano}. Since 
the subtracted $g_{pm}(t)$ behaves as ${\cal O}(\frac{1}{m^3})$, 
only the massive modes around $m=3H/2$ are relevant. 
Such massive modes are able to penetrate easily as    
$lH$ becomes large, because $lH$ becomes large, $\tilde{y_0}$ becomes 
large and the 'depth' of the volcano becomes shallow.

The difference between our model and RS model is the asymptotic value.
In RS model, volcano potential asymptotically approaches to 0.
So when $lH$ becomes large, 'volcano' disappears completely and it 
breaks the localization of 0-mode. In the end, Newtonian gravity on 
the brane breaks down. In contrast, there continues to exist the 
volcano in our model because of the non-zero asymptotic value.
Numerical calculation and (\ref{eq:asymptotic}) shows that the 
contribution of massive modes to the spectrum 
is enough small compared with 0-mode even at large $lH$.
As mentioned before, $|f_m(y_0)|^2/|f_0(y_0)|^2 \sim 0.080$ at $m=1.9$. 
Thus the ratio of massive modes to 0-mode at large $lH$ and 
$m=1.9$ becomes about 0.017. This is the maximum value of the ratio.
If we integrate about $m$ at large $lH$, the ratio becomes 0.03, 
namely, the contribution of 
0-mode is more than 30 times larger than that of massive modes.

At last, we refer to 0-mode. 0-mode is different from the mode 
which appears in the usual four-dimensional theory by some coefficients. 
So the amplitude of 0-mode has changed, but 
$\mbox{\boldmath $p$}$-dependence of 0-mode does not change at all. 

\section{Conclusion}
We have proposed an inflationary scenario in the brane world, i.e., 
inflation driven by the bulk inflation. The fluctuations of the bulk 
inflation has been also calculated. 
The contribution of massive modes is too small to change the spectrum 
at super horizon scale. So we can conclude that massive modes 
can not be observed even if $lH$ is very large. 

Here we refer to the density perturbation. In the conventional 
(i.e., four-dimensional) inflationary scenario, the relation between 
the density perturbation and $\delta\phi_4$ is well known.  
In contrast, the scalar field living in the bulk has both $t$ and 
$y$-dependence. There may exist the non-trivial background solution 
due to this $y$-dependence. If so, the density perturbation in this 
case may change drastically compared with that of the four-dimensional case. 
But if the mass of the scalar field is enough small, we can use the 
slow-roll approximation and we expect to get the almost same result 
as the four-dimensional case. 
 
Now, there is a point we do not refer in this research. 
That is the evolution after the inflation in our context. 
Cosmological evolution has been investigated in \cite{KS2}-\cite{BWCP4}.
Considering those results, there is the possibility that massive modes 
start dominating in the evolution after the inflation even if 
massive modes do not dominate at the end of the inflation.   
So we can say that concerning the initial spectrum, the brane-world 
model has survived and our scenario is one possible candidate where 
'inflation' and 'brane' are compatible with each other.


\section*{Acknowledgements}
We would like to thank M.Sasaki and Y.Himemoto for discussions. 
The authors are also grateful to H.Kodama for useful comments. 
The work of K.K. was supported by JSPS Research Fellowships for
Young Scientist No.4687.



\end{document}